\documentclass[aps,twocolumn,amsmath,amssymb,superscriptaddress,showpacs]{revtex4-2}%
\usepackage{graphicx}
\usepackage{bm}
\usepackage{float}
\usepackage{subfig}

\usepackage{xcolor}
\usepackage{braket}
\usepackage{tabularx}
\usepackage{comment}
\usepackage{afterpage}
\usepackage{placeins}
\usepackage{booktabs}
\usepackage{multirow}
\usepackage{array}
\usepackage{setspace}
\graphicspath{{Figs/}}
\usepackage{hhline}
\usepackage{xfrac}
\usepackage{mathtools}
\usepackage{listings}
\usepackage[normalem]{ulem}
\usepackage{titlesec}
\usepackage{amsfonts}
\usepackage[version=4]{mhchem}
\renewcommand\thesubsection{\Alph{subsection}}
\usepackage{epstopdf}
\catcode`@11
\def\seceqaa{\@addtoreset{equation}{section}
principles\def\theequation{A\arabic{equation}}}
\def\seceqbb{\@addtoreset{equation}{section}
\def\theequation{B\arabic{equation}}}
\def\seceqcc{\@addtoreset{equation}{section}
\def\theequation{C\arabic{equation}}}
\def\seceqdd{\@addtoreset{equation}{section}
\def\theequation{D\arabic{equation}}}
\def\seceqee{\@addtoreset{equation}{section}
\def\theequation{E\arabic{equation}}}
\def\seceqff{\@addtoreset{equation}{section}
\def\theequation{F\arabic{equation}}}
\def\seceqgg{\@addtoreset{equation}{section}
\def\theequation{G\arabic{equation}}}
\def\seceqhh{\@addtoreset{equation}{section}
\def\theequation{H\arabic{equation}}}
\catcode`@11

\newcommand{\mz}[1]{{#1}}
\newcommand{\zc}[1]{{#1}}
\newcommand{\zz}[1]{{#1}}
\newcommand{\mc}[1]{{#1}}

\begin{document}
\title{Influence of interactions on the chiral effect in $1D$ Dirac semimetal}
\author{Maksim Ulybyshev}
\address{ \scriptsize Institut f\"ur Theoretische Physik und Astrophysik, Universit\"at W\"urzburg, 97074 W\"urzburg, Germany}

\author{ Mikhail Zubkov}
\address{ \scriptsize Department of Physics, Ariel University, Ariel-40700, Israel}

\date{\today}

\begin{abstract}

\noindent

We consider the 1D Su--Schrieffer--Heeger (SSH) model. It was recently shown that, for the noninteracting model in its Dirac semimetal phase, the linear response of the axial charge density to an external electric field is proportional to the electrical conductivity in the presence of finite dissipation, with the proportionality factor determined by the coupling constants. This relation may be viewed as a manifestation of the chiral effect, which is a dimensional reduction of the 3D chiral magnetic effect. In the present work, we investigate the same model in the presence of two versions of local Hubbard-type interactions using numerical Quantum Monte Carlo simulations. We \zz{find, within the numerical resolution and for the parameters studied,} that, even in the regime where sufficiently strong interactions drive the system into a Mott insulating phase, the proportionality between the induced axial charge density and the electrical conductivity remains unchanged. This result indicates that the chiral effect is not renormalized by local Hubbard interactions.

\end{abstract}

\maketitle

\section{Introduction}

The Su--Schrieffer--Heeger (SSH) model has been proposed originally for the description of  polyacetylene \cite{PhysRevLett.42.1698} and later adopted for the description of optical lattices and other condensed matter systems \cite{PhysRevB.108.195103,PhysRevA.100.012112, PhysRevB.96.125418,qin2023one}. In \cite{meier2016observation} the experimental investigation of such a 1D system was reported. This one-dimensional model contains the Dirac semimetal phase, in which the chiral anomaly plays an important role.   

The chiral anomaly is reflected in the non-conservation of the axial current \cite{ZUMINO1984477}, despite its exact conservation in classical mechanics. It occurs because chiral symmetry is broken by quantum mechanical effects. 
The chiral anomaly is even more important in the three - dimensional materials 
like Weyl and Dirac semimetals, where it produces particle - hole pairs in the presence of parallel electric and magnetic fields \cite{PhysRevB.108.195103, PhysRevB.27.6083, PhysRevLett.64.1812, Lohse2016, Nakajima2016}. The observation of chiral anomaly in condensed matter systems is known for more than thirty years (see, for example, \cite{bevan1997momentum}).

In \(1+1\)-dimensional systems with gapless fermions, chiral anomaly produces chiral imbalance. In turn, this disbalance results in the appearance of electric current. This may be considered as the {\it chiral effect}, which is a one-dimensional cousin of the chiral magnetic effect (CME) that takes place in the three dimensional materials. As a result, the electric conductivity is governed not only by conventional scattering mechanisms but also by the chiral effect \cite{PhysRevB.27.6083, PhysRevLett.64.1812, Lohse2016, Nakajima2016}. 

The 1D Dirac semimetal appears as a dimensionally reduced 3D Dirac semimetal in the strong external magnetic field. The lowest Landau level dominates the electronic transport, which results in the effectively one-dimensional dynamics  \cite{Gusynin1999pq}. In \cite{abramchuk2024magnetoconductivity,ABRAMCHUK2026113374} it has been shown that in the presence of dissipation the chiral density and electric current are equal (up to a dimensional coefficient composed of the coupling constants). This observation confirms the presence of chiral magnetic effect {\cite{CMEZrTe5,Kharzeev2017}}. As a result, negative magnetoresistance in Dirac semimetals is due to the chiral magnetic effect (CME) at least at zero temperature and sufficiently strong magnetic field, in correspondence with  \cite{CMEZrTe5,Kharzeev2017}. In \cite{abramchuk2024magnetoconductivity,ABRAMCHUK2026113374}  the authors applied Keldysh technique unified with the specific version of Wigner-Weyl calculus \cite{chernodub2017scale,zhang2020influence}. The results on the CME in Dirac semimetals remind us about the other non-dissipative transport phenomenon - the chiral separation effect   \cite{zubkov2023effect}, which is governed by a topological invariant composed of the Green  function, and also the chiral vortical effect  \cite{abramchuk2018anatomy}. Presumably, the same topological invariant is intimately connected to the CME. This analogy extends the approach of \cite{Volovik2003a} to connection between condensed matter systems and high energy physics  \cite{zubkov2018momentum,zubkov2012momentum,volovik2017standard,volovik2013nambu,volovik2015scalar} based on the common topological invariants. Momentum space topological invariants have been discussed, for example, in \cite{zubkov2012momentum} \cite{zubkov2017topology}. These topological invariants are similar to those responsible for the quantum Hall effect  \cite{zhang2019hall,selch2025non}. General relation between two-dimensional condensed matter systems and relativistic quantum field theory is illustrated by physics of graphene  \cite{katsnelson2013euler}. 

It is worth mentioning that the CME exists also in quark matter, where the interactions are essentially non-perturbative, and various topological defects dominate dynamics  \cite{bakker1999central,bakker2005standard}. 

A disadvantage of calculations reported in  \cite{abramchuk2024magnetoconductivity,ABRAMCHUK2026113374} is that they were performed within the effective continuum theory instead of the more realistic tight-binding models. The step towards consideration of CME in the tight-binding models of Dirac/Weyl semimetals has been performed in \cite{bohra2025relation}, where the authors concentrate on the SSH model in the Dirac semimetal phase. It can be \mc{considered} as a result of a dimensional reduction of the 3D Dirac semimetal. The \mc{calculation} similar to that of \cite{abramchuk2024magnetoconductivity,ABRAMCHUK2026113374} showed that the chiral effect \mc{still} takes place, and the chiral disbalance, resulted from the chiral anomaly, is the source of conductivity. In the present paper we continue this line of research and extend the consideration of \cite{bohra2025relation} to the SSH model with Hubbard interactions between electrons.

\section{Chiral effect in 1D Dirac semimetals}
\label{SectIntro}

Several non - identical approaches  argue that magneto-resistance in Weyl and Dirac semimetals, as well as in quark matter, is related to the chiral magnetic effect (see, for example,  \cite{Fukushima2008,Kharzeev2017,Stephanov2012ki,Burkov2014,Son2013,Gorbar2016,Gao2012,Lin2019,Hattori2016lqx,Huang2017}). 
Among these approaches kinetic theory is worth to be mentioned  \cite{Son_2012,Stephanov2012ki,Gao2012,PhysRevB.96.235134,sekine2021axion},  as well as the quasiclassical approximation \cite{Burkov2014,Son2013,Gorbar2016} 
and other methods  \cite{Hattori2016lqx,Lin2019}. 
In the $^3$He-A superfluid (with electrically neutral excitations) there exists a special type of chiral magnetic effect resulted from the  emergent gauge field \cite{volovik2017chiral}.

In strong magnetic fields we can use Landau level representation, in which the lowest Landau level dominates. Correspondingly, motion in plane orthogonal to the magnetic field is excluded from consideration, and the dynamics is reduced to that of the motion along the direction of magnetic field, i.e. we arrive at the effective 1D model. This way in \cite{abramchuk2024magnetoconductivity} the CME was investigated in the effective continuum model. In particular, the Wigner-Weyl calculus was used \cite{chernodub2017scale,zhang2020influence,suleymanov2019wigner,suleymanov2019wigner}.

Let us repeat then the basic discussion of CME for the $1D$ Dirac semimetal. We arrive at the following basic assumptions:

\begin{enumerate}
	
	\item
	
	In the presence of electric field  $\vec E$ (directed along the only coordinate axis) the chiral anomaly pumps pairs (left-handed electron and right-handed hole or vice versa) from the Dirac sea of occupied energy levels. A version of the calculation is represented in \zc{Ref.~\cite{bohra2025relation}.} Qualitatively it is clear that this process results in the appearance of chiral imbalance (i.e. the difference between the densities $\rho_R$, $\rho_L$ of left-handed and right-handed fermion excitations). Naive calculations result in the following expression for the density of chiral charge:
	\begin{equation}
		\rho_5 = \rho_R - \rho_L = N_f\frac{\zc{e}{E} }{{\pi}\hbar}\,\, \tau_5, \label{EqRho5CME}
	\end{equation}
	where $N_f$ is the number of Dirac fermions in the given system, while $\tau_5$ is a relaxation time. 
	{\it Here we use SI units.}
	
	Eq. (\ref{EqRho5CME}) may be obtained as a solution of  phenomenological kinetic equation that accounts for the  chirality relaxation. In turn, such an  equation is to come as an  approximation to the precise kinetic theory (Keldysh--Schwinger technique).
	
	\item 
	
	The next general assumption is that the chiral imbalance is related to the notion of chiral chemical potential $\mu_5 = (\mu_R - \mu_L)/2$ in the way similar to relation of the total number of electrons and the ordinary chemical potential. Namely, assuming the presence of the corresponding steady state we have 
	\begin{equation}
		\rho_5 \approx {N_f}\frac{\mu_5}{\pi {v_F\hbar}}. \label{EqMu5CME}
	\end{equation}

	\item
	
	Next, we formulate the one - dimensional cousin of CME, i.e. {\it the chiral effect} to express electric current through chiral chemical potential (this formula is obtained from the  standard expression for the electric current of the CME  dividing it by the density  of states $\zc{e}B/(2\pi \hbar)$ in the plane orthogonal to magnetic field) 
	
	\begin{equation}
		{ j} = {N_f}\zc{\frac{e}{\pi\hbar}} \mu_5 \label{CME}
	\end{equation}
	and the value of electric conductivity is given by	
	{\begin{equation}
			\sigma^{zz}_{CME} =  N_f \zc{\frac{e^2 v_F}{{\pi}\hbar}}  \tau_5 \label{EqCond2CME}
	\end{equation}}
	which may also be obtained via dimensional reduction of Kubo formula from \cite{gorbar2014chiral},{\cite{Lu_2015,Li_2023}}.
	
\end{enumerate}   

One can see comparing Eqs. (\ref{EqMu5CME}) and (\ref{CME}) that electric current and chiral density coincide up to the factor $\zc{e v_F}$:
\begin{equation}
	j = \zc{e} v_F \rho_5\label{CME2}
\end{equation}
 This equality excludes from consideration \mc{the axial chemical potential} $\mu_5$ that does not have a direct meaning in the theory out of equilibrium. Eq. (\ref{CME2}), therefore, represents an alternative formulation of 1D chiral effect that is a dimensionally reduced CME.

\section{The Su-Schrieffer-Heeger (SSH) Model}
The Su-Schrieffer-Heeger (SSH) model describes a dimerized chain \cite{PhysRevLett.42.1698, PhysRevA.95.061601, PhysRevB.96.125418, PhysRevA.100.012112}. 
The  Hamiltonian of the non - interacting SSH model is given by
\begin{eqnarray}
    \hat H_0 &=& \sum_{i,\sigma} \Bigl( {\hat c}_{iA\sigma}^\dagger J_0 {\hat c}_{iB\sigma} + {\hat c}_{(i+1)A\sigma}^\dagger J_1 {\hat c}_{iB\sigma}\nonumber\\&& + {\hat c}_{(i+1)B\sigma}^\dagger J_2 {\hat c}_{iA\sigma} + \text{H.c.} \Bigr) \nonumber\\ &=&\sum_{i,\sigma} \Bigl( {\hat \psi}_{i,\sigma}^\dagger T_0 {\hat \psi}_{i,\sigma} + ({\hat \psi}_{i+1,\sigma}^\dagger T_1 {\hat \psi}_{i,\sigma} + \text{H.c.}) \Bigr),\label{HT}
\end{eqnarray}
\mz{with $\psi_{i,\sigma} = (c_{iA\sigma},c_{iB\sigma})^T $}. In the present paper we consider the two versions of the model: with the spinful electrons $\sigma = \uparrow, \downarrow$ ({\bf Model 1}), and with the spinless electrons $\sigma = 0$ ({\bf Model 2}). Matrices entering Eq. (\ref{HT}) have the form 
\begin{equation}
T_0 = \begin{pmatrix} 0 & J_0 \\ J_0 & 0 \end{pmatrix}, \quad T_1 = \begin{pmatrix} 0 & J_1 \\ J_2 & 0 \end{pmatrix},
\end{equation}
\zz{For fixed $J_1+J_2=J_0=1$,} the hopping matrices are characterized by the dimerization parameter $2\Delta = J_1 - J_2$.
For $\Delta = 0$ the dispersion (close to its minimum) is quadratic in momentum. Finite value $\Delta \neq 0$ changes this pattern. There exists a choice of parameters corresponding to the single gapless Dirac cone. In particular, we choose  
	\begin{equation}
		J_0 = 1, \quad J_1 = 1, \quad J_2 = 0,\label{par}
\end{equation}
From now on in the present paper we use for simplicity of expressions \zz{units with $\hbar=1$ and measure energies in units of $J_0$; thus, the displayed values of $J_i$ are dimensionless ratios.}

Besides, \zz{for the considered parameters and a unit lattice spacing, the Fermi velocity is $v_F=1$ in these units.} Also for simplicity in the following we absorb electric charge $e$ into the definition of electric field, while electric current is measured in units of $e$. As a result factor $\zc{e v_F}$ disappears from Eq. (\ref{CME2}).
	
\zz{Restoring dimensions, the Hamiltonian of Eq. (\ref{HT}) is multiplied by the chosen energy unit.}
The Hamiltonian of this model in momentum space is 
\begin{equation}
	\mathcal{H}(k) = (1 + \cos(k)) \sigma_1 + \sin(k) \sigma_2.
\end{equation}
 \mc{The} SSH model for the values of parameters of Eq. (\ref{par}) represents the simplest one - dimensional Wilson fermions with vanishing mass, well known in relativistic lattice field theory. In \cite{bohra2025relation} it has been proven that if we supplement the non - interacting model with the finite dissipation rate $\epsilon$ (which captures somehow both effects of disorder and interactions with thermal bath of phonons), then Eq. (\ref{CME2}) indeed holds, while $\tau_5$ is given by $\hbar/(2 \epsilon)$.

The similar calculation has been performed for the 3+1D system in the presence of strong external magnetic field \cite{abramchuk2024magnetoconductivity}, which makes the system effectively 1+1 dimensional (including the evaluation of the dissipation rate due to disorder and interaction with phonons). However, in \cite{abramchuk2024magnetoconductivity} the effective continuum low energy effective field theory with emergent relativistic invariance was considered, while in \cite{bohra2025relation} we investigated a tight - binding model describing the real lattice systems. 

The purpose of the present paper is to check the validity of Eq. (\ref{CME2}) in the presence of interactions. We consider the Hubbard interactions and recover the validity of Eq. (\ref{CME2}) for the considered values of parameters.

\section{SSH model with Hubbard interactions }

\subsection{Simulation setup and observables}

In {\bf Model 1} the standard Hubbard interaction term is added to the Hamiltonian
 
\begin{equation}
	\hat H_{Hub.}= U \sum_{i,\mu} (\hat {n}_{i, \mu,\uparrow} - \frac{1}{2} ) (\hat {n}_{i, \mu,\downarrow} - \frac{1}{2} ),
\end{equation}
where $\mu = A,B$, $\hat {n}_{i,\mu,  \sigma}=\hat c^\dag_{i \mu \sigma} \hat c_{i \mu \sigma}$. 

In the spinless {\bf Model 2} we add the following interaction term
\begin{equation}
	\hat H_{Hub.}= U \sum_{i} ( \hat {n}_{i, A} - \frac{1}{2} ) ( \hat {n}_{i, B}- \frac{1}{2} ) ,
\end{equation}
where $\hat {n}_{i,  A,B}=\hat c^\dag_{i  A, B} \hat c_{i  A, B} $.

In both cases the standard Hubbard-Stratonovich decomposition of the interaction term is possible using discrete auxiliary field \cite{Blankenbecler81,White89}. Subsequently, we employ the BSS-QMC algorithm \cite{Assaad08_rev} encoded in \cite{superlat_github} for finite temperature auxiliary field QMC simulations of the full interacting Hamiltonian.

We investigate the two basic quantities: static response $\lambda(0)$ of axial density 
\begin{equation}
\hat \rho_5 = \zc{\hat {\psi}^\dagger} \sigma^2 \hat \psi
\end{equation}
 to external electric field, and the static electric conductivity $\sigma(0)$. \zz{Here $\hat J$ below denotes the spatially averaged electric-current operator obtained by differentiating the Peierls-substituted Hamiltonian with respect to the vector potential.} According to the fluctuation - dissipation theorem these quantities are expressed through the spectral densities of the correlators: 
\begin{eqnarray}
	\langle \hat J(0) \hat J(\tau)\rangle = \int_0^\infty \frac{\omega}{\pi} \frac{\cosh (\omega (\tau-\beta/2) )}{\sinh(\omega\beta/2)} \sigma (\omega) d\omega
\end{eqnarray}
and 
\begin{eqnarray}
	\langle \hat \rho_5(0) \hat J(\tau)\rangle = \int_0^\infty \frac{\omega}{\pi} \frac{\cosh (\omega (\tau-\beta/2) )}{\sinh(\omega\beta/2)} \lambda (\omega) d\omega
\end{eqnarray}
By $\beta$ we denote here the inverse temperature in lattice units.

In addition to that, we also compute the spectral density $A(\omega,k)$ of the electron Green function in order to check the stability of the Fermi point to the introduction of interaction. \zz{Using particle--hole symmetry,} this spectral density is defined through the relation
\begin{eqnarray}
	\langle \mc{\hat {\psi}^\dagger} (0,k) \hat \psi(\tau, k)\rangle = \int_0^\infty \frac{1}{\pi} \frac{\cosh (\omega (\tau-\beta/2) )}{\cosh(\omega\beta/2)} A (\omega,k) d\omega
\end{eqnarray}

Spectral functions are reconstructed from the respective Euclidean time correlators using stochastic analytical continuation method 
\cite{Beach04a,Sandvik98,SHAO20231}.

\subsection{The known facts about the considered models}

  {\bf Model 2} contains one pair of spinless gapless electrons:  right - handed and left - handed ones. Both are located close to $p = \pi$ in momentum space. Noninteracting version of {\bf Model 1} is the doubled {\bf Model 2} with spin degeneracy. Standard Hubbard interaction is added to {\bf Model 1}, and at zero temperature this model is exactly solvable (see, for example, \cite{DEGUCHI2000197} and references therein). In particular, the \zz{small-$U$ asymptotic form of the energy gap} is known: 
	\begin{equation}
		\Delta \zz{\simeq} \frac{8 \sqrt{U}}{\pi}e^{-\frac{2\pi}U}\zz{,\qquad U\ll 1} 
		\label{exp_gap}
	\end{equation}
	For the systems at finite temperature, this gap can not be even detected in numerical simulations \zz{at sufficiently weak $U$}. For example, at temperatures   $T = 0.05$ and $T = 0.25$ ($\beta = 20$ and $\beta = 4$) the system behaves essentially as Dirac semimetal, and we can use it \zc{to} check the chiral effect \zc{(Eq. (\ref{CME2}))}. It is worth mentioning that while the given model is formally exactly solvable, there is no precise analytical result for the considered quantities at finite temperatures. Our results demonstrate that the equality $\rho_5 = j$ remains robust with respect to the introduction of interactions at the considered values of temperature, which reveals the essence of the chiral effect in the interacting systems.
	
	{\bf Model 2} contains spinless electrons with \zz{intracell nearest-neighbour} density-density interactions. It is not exactly solvable, and because of the absence of the second flavor of electrons the gap is not suppressed exponentially at small $U$. Instead, it grows linearly in $U$, which is observed via numerical simulations performed at the values of temperature $T = 0.05$ and $0.25$. We detect the gap looking at the location of the peak in the spectral function of electron propagator at the former Dirac point. Finite temperature does not make this gap invisible unlike the case of {\bf Model 1}, however, at sufficiently small interaction, the finite temperature fluctuations (in $T = 0.25$ case) result in the appearance of thermal conducting electrons. Correspondingly, the system still features noticeable \zz{electrical conductivity due to thermally activated carriers}, and we again are able to check the validity of the chiral effect \zc{(Eq. (\ref{CME2}))}. Our results demonstrate here that as well as in {\bf Model 1} \zz{the linear-response coefficients of the electric current and chiral density coincide in the $\omega\to0$ limit}.

\section{Numerical results}

\subsection{Model 1}

\begin{figure}[ht]
	\center{\includegraphics[angle=0,width=0.9\linewidth]{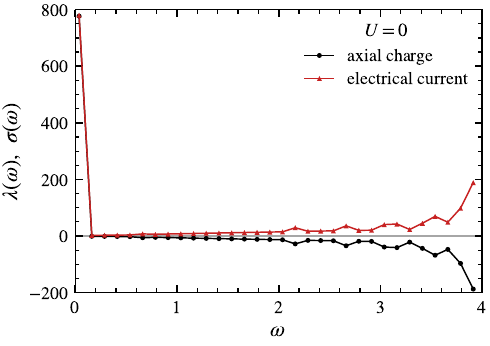}}
	\caption{Model 1. Spectral functions $\lambda(\omega)$ and $\sigma(\omega)$ in the absence of interactions.
		\label{figU0}
	}
\end{figure}

\begin{figure}[ht]
	\center{\includegraphics[width=0.9\linewidth]{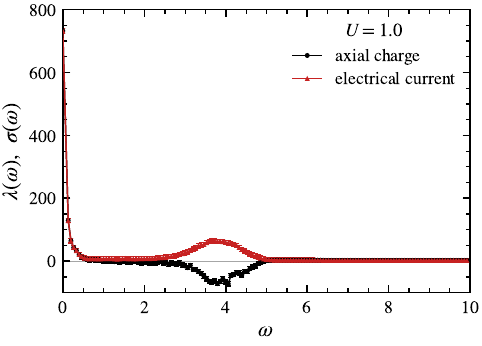}}
	\caption{Model 1. Spectral functions $\lambda(\omega)$ and $\sigma(\omega)$ in the presence of interactions at $U = 1.0$.
		\label{figU1}
	}
\end{figure}

\begin{figure}[ht]
	\center{\includegraphics[width=0.9\linewidth]{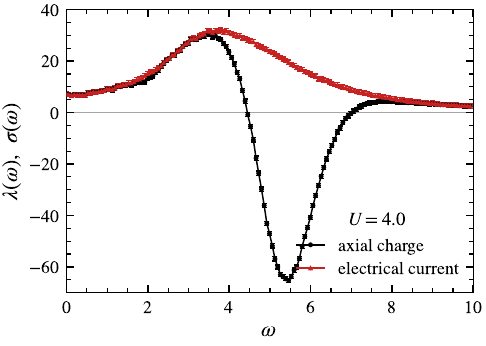}}
	\caption{Model 1. Spectral functions $\lambda(\omega)$ and $\sigma(\omega)$ in the presence of interactions at $U = 4.0$.
		\label{figU4}
	}
\end{figure}

In order to calculate electric conductivity and axial charge density response to electric field we fix the inverse temperature to be 
$\beta = 4.0$, the number of lattice sites is $100$. The values of parameter $U$ are chosen to be equal to $0, 1.0, 4.0$. The non-interacting results are presented on Fig. \ref{figU0}: since the spectral function on finite lattice is just a set of delta-functions, we averaged them over intervals in between of points shown in the plot. One can see the peak at zero frequency, which corresponds to single delta-function exactly at $\omega = 0$. Coefficient in front of this delta-function is the same for $\lambda(\omega)$ and $\sigma(\omega)$. In general we see, that although the finite-frequency responses for axial charge and current are different, $\lambda(\omega) \rightarrow \sigma(\omega)$ at small frequencies, as expected. Delta-function exactly at zero gives us an infinite DC conductivity as it should be in the absence of dissipation. 
The dissipation appears once we introduce Hubbard interaction $U$. One can see the results in Fig. \ref{figU1} for $U=1.0$: Drude peak  acquires finite width. Finally, the interactions effects are even more pronounced at $U= 4.0$ - see Fig. \ref{figU4}, where Drude peak is leveled out in both  $\lambda(\omega)$ and $\sigma(\omega)$. 

\begin{figure}[ht]
	\center{\includegraphics[width=0.9\linewidth]{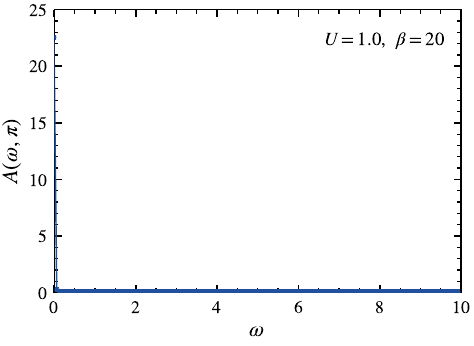}}
	\caption{Model 1. Spectral function $A(\omega,\pi)$ at the Fermi point position $k = \pi$ in the presence of interactions at $U = 1.0$ at $\beta = 20$.
		\label{figA1}
	}
\end{figure}

\begin{figure}[ht]
	\center{\includegraphics[width=0.9\linewidth]{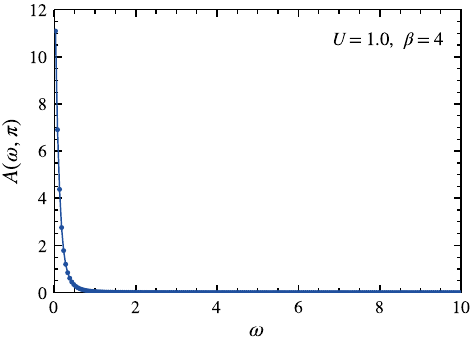}}
	\caption{Model 1. Spectral function $A(\omega,\pi)$ at the Fermi point position $k = \pi$ in the presence of interactions at $U = 1.0$ and $\beta = 4$ .
		\label{figA1_}
	}
\end{figure}

\begin{figure}[ht]
	\center{\includegraphics[width=0.9\linewidth]{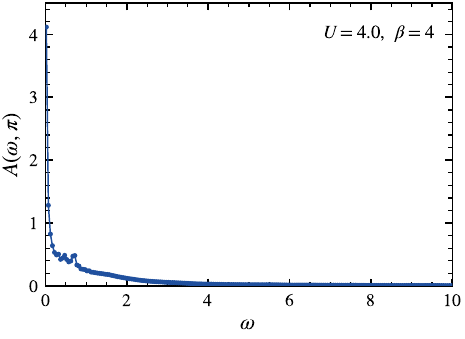}}
	\caption{Model 1. Spectral function $A(\omega,\pi)$ at the Fermi point position $k = \pi$ in the presence of interactions at $U = 4.0$ and $\beta = 4$ .
		\label{figA4_}
	}
\end{figure}

\begin{figure}[ht]
	\center{\includegraphics[width=0.9\linewidth]{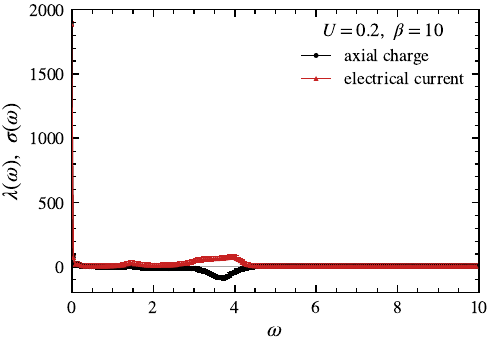}}
	\caption{Model 1. Spectral functions $\lambda(\omega)$ and $\sigma(\omega)$ in the presence of interactions at $U = 0.2$ and $\beta = 10$.
		\label{figU02b10}
	}
\end{figure}

The single-electron spectral density  $A(\omega,\pi)$ at $\beta = 20$ is shown on Fig. \ref{figA1} for $U = 1.0$ (Without interactions the Fermi point is situated at $k = \pi$) One can see that there is no signature of the gap opening. The same can be seen for $\beta = 4$ on Fig. \ref{figA1_} for $U=1.0$ and Fig.  \ref{figA4_} for $U=4.0$. At this temperature the peak of the spectral function is still concentrated at $\omega  = 0$ even for  relatively large value of $U$, which reflects the presence of thermal charge carriers at finite temperatures \zz{despite the interaction-induced gap; at weak coupling this gap is exponentially small} \zc{(Eq. (\ref{exp_gap}))}. The spectral densities of correlators $\lambda$ and $\sigma$ in the low temperature limit are shown in Fig. \ref{figU02b10} for $U=0.2$ and $\beta = 10$. One can see that the two spectral functions still coincide at $\omega \to 0$ for this model even in the low temperature limit.

\subsection{Model 2}

As in the case of Model 1 we fix the inverse temperature $\beta = 4$ for the calculation of response to electric field of $\rho_5$ and $j$. The values of parameter $U$ are chosen to be equal to $\zc{0.01}, 1.0, 4.0$. The non-interacting result repeat the ones of Model 1 (recall that without interaction Model 1 is equal to the doubled Model 2). Thus we again obtain the delta-functional peak at $\omega = 0$, and here also $\lambda(\omega) = \sigma(\omega)$ in the $\omega \rightarrow 0$ limit.

\begin{figure}[ht]
	\center{\includegraphics[width=0.9\linewidth]{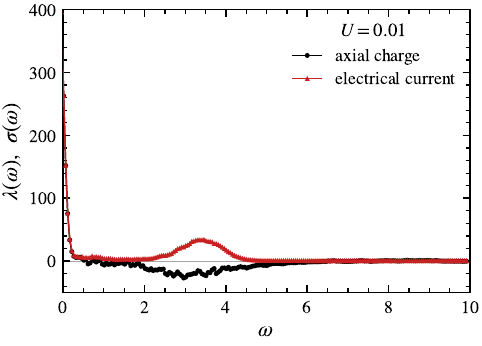}}
	\caption{Model 2. Spectral functions $\lambda(\omega)$ and $\sigma(\omega)$ in the presence of interactions at $U = 0.01$.
		\label{figU422}
	}
\end{figure}

\begin{figure}[ht]
	\center{\includegraphics[width=0.9\linewidth]{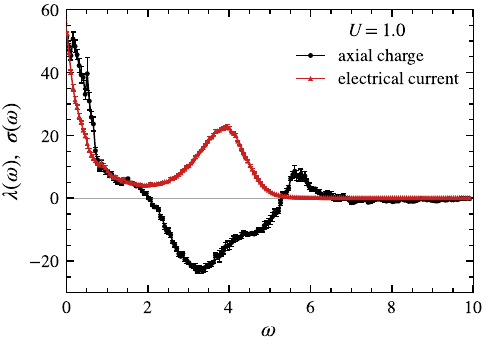}}
	\caption{Model 2. Spectral functions  $\lambda(\omega)$ and $\sigma(\omega)$ in the presence of interactions at $U = 1.0$.
		\label{figU12}
	}
\end{figure}

\begin{figure}[ht]
	\center{\includegraphics[width=0.9\linewidth]{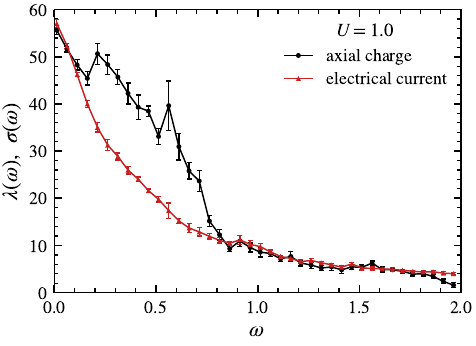}}
	\caption{Model 2. Spectral functions $\lambda(\omega)$ and $\sigma(\omega)$ in the presence of interactions at $U = 1.0$ (same plot as Fig. \ref{figU12}, but only low frequency parts of the spectral functions are plotted).
		\label{figU12_}
	}
\end{figure}

Now we introduce current dissipation by including nonzero Hubbard interaction $U$. First, at very small interaction $U=0.01$, we see no real changes in comparison to the free case (see Fig. \ref{figU422}). At moderate interaction $U=1.0$, the Drude peak develops larger width reflecting an \zz{increased scattering rate (shorter relaxation time)}. One can see the results in Figs. \ref{figU12} and \ref{figU12_}. 
Finally, at $U= 4.0$ the Drude peak is completely washed away (see Figs. \ref{figU42} and \ref{figU42_}) due to the gap being larger than temperature (see below).
\begin{figure}[ht]
	\center{\includegraphics[width=0.9\linewidth]{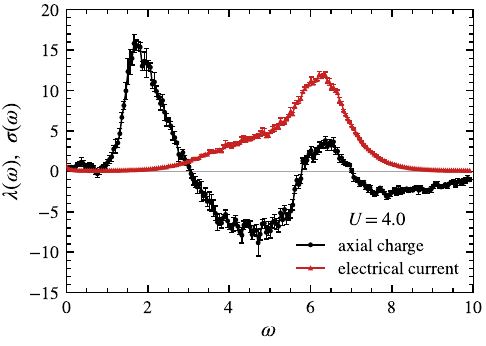}}
	\caption{Model 2. Spectral functions $\lambda(\omega)$ and $\sigma(\omega)$ in the presence of interactions at $U = 4.0$.
		\label{figU42}
	}
\end{figure}
\begin{figure}[ht]
	\center{\includegraphics[width=0.9\linewidth]{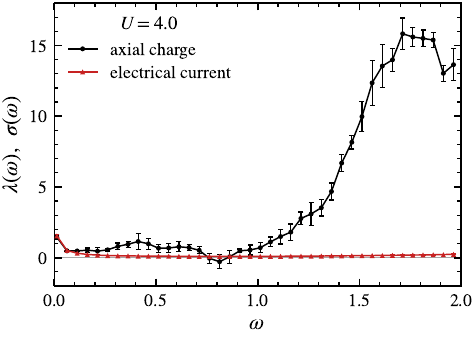}}
	\caption{Model 2. Spectral functions $\lambda(\omega)$ and $\sigma(\omega)$ in the presence of interactions at $\zc{U = 4.0}$  (same plot as Fig. \ref{figU42}, but only low frequency parts of the spectral functions are plotted).
		\label{figU42_}
	}
\end{figure}
\begin{figure}[ht]
	\center{\includegraphics[width=0.9\linewidth]{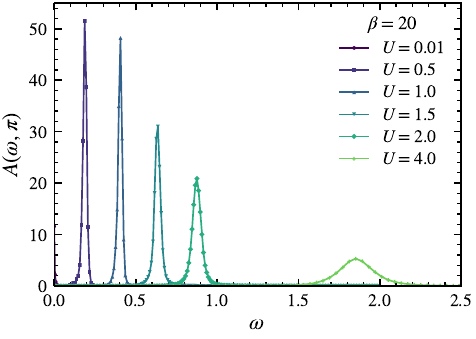}}
	\caption{Model 2. Spectral function $A(\omega,\pi)$ at the Fermi point position $k = \pi$ in the presence of interactions for $\beta = 20$ as functions of $U$.
		\label{figA12}
	}
\end{figure}

The single-electron spectral density $A(\omega,\pi)$ is shown on Fig. \ref{figA12} for $\beta = 20$  (again, in absence of interactions, the Fermi point is situated at $k = \pi$). One can see that the gap is vanishing for vanishing $U$ and grows almost linearly in $U$. This result means that the Fermi point is again as in the case of Model 1 not robust to the introduction of interactions, but this time the gap is actually visible even for relatively small interaction $U=0.5$. However, the gap is relatively small at $U=1.0$: it is comparable to temperature in $\beta=4$ simulations, thus we still could see the residual \zz{electrical conductivity from thermally activated carriers} in Figs. \ref{figU12} and \ref{figU12_}. At $U=4.0$ the gap is roughly 1.8, thus substantially exceeding the temperature. Hence \zc{the} disappearance of the thermal \zc{charge} carriers, and the absence of the Drude peak on the Figs. \ref{figU42} and \ref{figU42_}.



\section{Conclusions and discussion}

The chiral effect in 1D systems appears as a dimensional reduction of chiral magnetic effect of 3D Dirac semimetals. At strong magnetic field the dynamics in the latter materials is reduced to just \zc{one} dimension, along the direction of magnetic field. 
	
In the previous work \cite{bohra2025relation} it has been shown using analytical methods that the chiral effect takes place in the Dirac semimetal phase of the SSH model and is realised as a relation \zc{(\ref{CME2})} between electric current and axial charge density (up to a dimensional factor). 

In the present work we observe with the help of numerically exact Quantum Monte Carlo simulations that the same relation holds in the presence of local Hubbard-type interactions. This means that, \zz{within the numerical accuracy and for the parameters studied,} the chiral effect is not renormalized by local density-density interactions. Moreover, we expect that it is not renormalized by any kinds of interactions, at least when such interactions are sufficiently weak. This \zz{suggests a possible topological origin of relation (\ref{CME2}) in the effectively gapless regime, within linear response}.   

In  \cite{bohra2025relation}  it has been proven analytically that  topological invariant
\begin{equation}
	 {\cal N } = \frac{1}{4\pi i {\bf L} } \int dx \int_{\cal C}  \mathrm{tr} \bigl( \gamma^5 G_W d Q_W \bigr) 
\label{topinv0}
\end{equation}
 enters expression for the chiral anomaly, when the interactions are neglected. Here $\bf L$ is the length of the system, $G_W$ is the Wigner transformed Green function, while $Q_W$ is the Weyl symbol of an operator inverse to the Green function. $\gamma^5 $ is chiraliry matrix, which in our case is equal to $\sigma^2$. Contour $\cal C$ consists of the two straight lines in momentum space $\omega = \pm 0$ and $\omega$ is the Matsubara frequency. One may expect that in the absence of an energy gap without interactions  
  the same quantity enters expression for the electric conductivity (and the response of axial density to electric field)  
\begin{equation}
	j = \rho_5 = {\cal N}\frac{E}{\pi} \tau,
\end{equation} 
Here $\tau$ is the relaxation time. For the considered SSH model the value of the topological invariant of Eq. (\ref{topinv0}) is equal to $2$ due to spin degeneracy. When the system contains a set of fermions, its value in the simplest case is equal to their number. However, in general case $\cal N$ may differ from this number  \cite{Volovik2003a}. 

As is well - known (and as we observe it in the present work), the gap is opened due to the interactions in both considered models. Hence, the quantity of Eq. (\ref{topinv0}) vanishes. Strictly speaking, in the presence of interactions we deal with Mott insulator instead of Dirac semimetal. In this situation \zz{the linear-response anomaly formula (\ref{EqRho5CME}) no longer applies, while nonperturbative pair creation may occur through a tunneling mechanism}. This situation has been considered, in particular, in \cite{oka2010dielectric}, where the following expression for the \zz{ground-state decay rate} per unit length has been obtained (using certain approximate methods):
\begin{equation}
	\frac{\Gamma}{\bf L} = -\frac{E}{2\pi}\, {\rm log}\Bigl( 1 - e^{-\pi \frac{F}{E}}\Bigr)\label{pair}
\end{equation}
Here $F$ is a parameter that depends on the energy gap $\Delta$, and vanishes when the latter vanishes. The \zz{scaling $F \propto \Delta^2/(\hbar v_F)$} is valid at weak electric fields with methods used in \cite{oka2010dielectric} giving a more complicated dependence of $F$ on $\Delta$ for strong $E$. The above expression for the \zz{ground-state decay rate} can be further improved at strong $E$ by a damping factor $a<1$ in order to match the other reported results (see references in \cite{oka2010dielectric}). 

\zz{If $\Gamma/{\bf L}$ is identified with the pair-production rate and each produced pair contributes two units of chiral charge, Eq. (\ref{pair}) gives}
\begin{equation}
	\rho_5 = \zz{-\frac{E}{\pi}\, {\rm log}\Bigl( 1 - e^{-\pi \frac{F}{E}}\Bigr) \tau_5}\label{rho5tun}
\end{equation}
where $\tau_5$ is the corresponding relaxation time that reflects the annihilation of the pairs. Correspondingly, the presence of chiral effect would mean that $j$ is given by the same expression. Such a study, however, would require the use of non-equilibrium interacting field theory. Therefore, such a study remains out of the scope of the present paper. However, we can study the linear response of both axial charge and current to electric field applying Kubo formulae to the Euclidean correlators obtained in thermal equilibrium on finite systems. 

Namely, in {\bf Model 1} the \zz{small-$U$ asymptotic form of the} energy gap depends on the Hubbard parameter $U$ as 
\begin{equation}
	\Delta \zz{\simeq} \frac{8 \sqrt{U}}{\pi}e^{-\frac{2\pi}U}\zz{,\qquad U\ll1}
\end{equation}
Due to the exponential suppression of this value at small $U$, the gap can not be detected at realistic temperatures, accessible for QMC simulations. Even our numerical results obtained at rather small temperature ($0.05$ in lattice energy units) confirm this expectation. It means that we can deal with this model as with a gapless one and the obtained results confirm the chiral effect at $T = 0.05$ for interacting 1D Dirac semimetal. This is not a surprise given the model is exactly solvable, and such an equality is expected based on the analytical results. However, the results become more nontrivial at larger temperatures (we considered also $T = 0.25$). In this case, no precise analytical result is known for electric conductivity (although certain exact analytical results for thermodynamic quantities are reported in literature). Rather unexpectedly we obtain that the chiral effect survives as well at this temperature, i.e. equality $\rho_5 = j$ still remains valid.   

We further considered {\bf Model 2}, which contains spinless electrons and \zz{intracell nearest-neighbour} Hubbard-type interactions. This model is not exactly solvable even at zero temperature. The energy gap is not suppressed exponentially at small $U$, and the gap can be directly observed in single-particle spectral function computed at finite $U$. However, at sufficiently large temperatures (already around $T = 0.25$ in lattice units) \zz{thermally activated carriers obscure the transport consequences of the energy gap}, and we can apply Kubo formulae to calculate electric conductivity as in the {\bf Model 1}. We again observe that the chiral effect survives, which results in equality $\rho_5 = j$. 

Thus we come to the conclusion that, at least in the considered systems and at least in the conditions that provide the "invisibility" of the energy gap the chiral disbalance remains the source of electric current in the presence of interactions. This is the $1D$ chiral effect that might be considered as the dimensional reduction of the $3D$ chiral magnetic effect.     
 
 \section*{Acknowledgements}
 
 We acknowledge the Ariel HPC Center at Ariel University for providing computing resources that have contributed to the research results reported within this paper.


\bibliographystyle{apsrev4-2}
\bibliography{../common_refs/refs,../common_refs/CMEwKeldysh_ext,../common_refs/CSE_MZ,../common_refs/references,../common_refs/biblio_corrected,../common_refs/wigner3,../common_refs/cross-ref,../common_refs/citationsZ,../common_refs/QMC_bib}

\begin{thebibliography}{61}%
\makeatletter
\providecommand \@ifxundefined [1]{%
 \@ifx{#1\undefined}
}%
\providecommand \@ifnum [1]{%
 \ifnum #1\expandafter \@firstoftwo
 \else \expandafter \@secondoftwo
 \fi
}%
\providecommand \@ifx [1]{%
 \ifx #1\expandafter \@firstoftwo
 \else \expandafter \@secondoftwo
 \fi
}%
\providecommand \natexlab [1]{#1}%
\providecommand \enquote  [1]{``#1''}%
\providecommand \bibnamefont  [1]{#1}%
\providecommand \bibfnamefont [1]{#1}%
\providecommand \citenamefont [1]{#1}%
\providecommand \href@noop [0]{\@secondoftwo}%
\providecommand \href [0]{\begingroup \@sanitize@url \@href}%
\providecommand \@href[1]{\@@startlink{#1}\@@href}%
\providecommand \@@href[1]{\endgroup#1\@@endlink}%
\providecommand \@sanitize@url [0]{\catcode `\\12\catcode `\$12\catcode
  `\&12\catcode `\#12\catcode `\^12\catcode `\_12\catcode `\%12\relax}%
\providecommand \@@startlink[1]{}%
\providecommand \@@endlink[0]{}%
\providecommand \url  [0]{\begingroup\@sanitize@url \@url }%
\providecommand \@url [1]{\endgroup\@href {#1}{\urlprefix }}%
\providecommand \urlprefix  [0]{URL }%
\providecommand \Eprint [0]{\href }%
\providecommand \doibase [0]{https://doi.org/}%
\providecommand \selectlanguage [0]{\@gobble}%
\providecommand \bibinfo  [0]{\@secondoftwo}%
\providecommand \bibfield  [0]{\@secondoftwo}%
\providecommand \translation [1]{[#1]}%
\providecommand \BibitemOpen [0]{}%
\providecommand \bibitemStop [0]{}%
\providecommand \bibitemNoStop [0]{.\EOS\space}%
\providecommand \EOS [0]{\spacefactor3000\relax}%
\providecommand \BibitemShut  [1]{\csname bibitem#1\endcsname}%
\let\auto@bib@innerbib\@empty
\bibitem [{\citenamefont {Su}\ \emph {et~al.}(1979)\citenamefont {Su},
  \citenamefont {Schrieffer},\ and\ \citenamefont
  {Heeger}}]{PhysRevLett.42.1698}%
  \BibitemOpen
  \bibfield  {author} {\bibinfo {author} {\bibfnamefont {W.~P.}\ \bibnamefont
  {Su}}, \bibinfo {author} {\bibfnamefont {J.~R.}\ \bibnamefont {Schrieffer}},\
  and\ \bibinfo {author} {\bibfnamefont {A.~J.}\ \bibnamefont {Heeger}},\
  }\href {https://doi.org/10.1103/PhysRevLett.42.1698} {\bibfield  {journal}
  {\bibinfo  {journal} {Phys. Rev. Lett.}\ }\textbf {\bibinfo {volume} {42}},\
  \bibinfo {pages} {1698} (\bibinfo {year} {1979})}\BibitemShut {NoStop}%
\bibitem [{\citenamefont {Qin}\ \emph {et~al.}(2023{\natexlab{a}})\citenamefont
  {Qin}, \citenamefont {Xu}, \citenamefont {Ning},\ and\ \citenamefont
  {Wang}}]{PhysRevB.108.195103}%
  \BibitemOpen
  \bibfield  {author} {\bibinfo {author} {\bibfnamefont {Z.}~\bibnamefont
  {Qin}}, \bibinfo {author} {\bibfnamefont {D.-H.}\ \bibnamefont {Xu}},
  \bibinfo {author} {\bibfnamefont {Z.}~\bibnamefont {Ning}},\ and\ \bibinfo
  {author} {\bibfnamefont {R.}~\bibnamefont {Wang}},\ }\href
  {https://doi.org/10.1103/PhysRevB.108.195103} {\bibfield  {journal} {\bibinfo
   {journal} {Phys. Rev. B}\ }\textbf {\bibinfo {volume} {108}},\ \bibinfo
  {pages} {195103} (\bibinfo {year} {2023}{\natexlab{a}})}\BibitemShut
  {NoStop}%
\bibitem [{\citenamefont {Du}\ \emph {et~al.}(2019)\citenamefont {Du},
  \citenamefont {Wu}, \citenamefont {Artoni},\ and\ \citenamefont
  {La~Rocca}}]{PhysRevA.100.012112}%
  \BibitemOpen
  \bibfield  {author} {\bibinfo {author} {\bibfnamefont {L.}~\bibnamefont
  {Du}}, \bibinfo {author} {\bibfnamefont {J.-H.}\ \bibnamefont {Wu}}, \bibinfo
  {author} {\bibfnamefont {M.}~\bibnamefont {Artoni}},\ and\ \bibinfo {author}
  {\bibfnamefont {G.~C.}\ \bibnamefont {La~Rocca}},\ }\href
  {https://doi.org/10.1103/PhysRevA.100.012112} {\bibfield  {journal} {\bibinfo
   {journal} {Phys. Rev. A}\ }\textbf {\bibinfo {volume} {100}},\ \bibinfo
  {pages} {012112} (\bibinfo {year} {2019})}\BibitemShut {NoStop}%
\bibitem [{\citenamefont {Li}\ \emph {et~al.}(2017)\citenamefont {Li},
  \citenamefont {Lin}, \citenamefont {Zhang},\ and\ \citenamefont
  {Song}}]{PhysRevB.96.125418}%
  \BibitemOpen
  \bibfield  {author} {\bibinfo {author} {\bibfnamefont {C.}~\bibnamefont
  {Li}}, \bibinfo {author} {\bibfnamefont {S.}~\bibnamefont {Lin}}, \bibinfo
  {author} {\bibfnamefont {G.}~\bibnamefont {Zhang}},\ and\ \bibinfo {author}
  {\bibfnamefont {Z.}~\bibnamefont {Song}},\ }\href
  {https://doi.org/10.1103/PhysRevB.96.125418} {\bibfield  {journal} {\bibinfo
  {journal} {Phys. Rev. B}\ }\textbf {\bibinfo {volume} {96}},\ \bibinfo
  {pages} {125418} (\bibinfo {year} {2017})}\BibitemShut {NoStop}%
\bibitem [{\citenamefont {Qin}\ \emph {et~al.}(2023{\natexlab{b}})\citenamefont
  {Qin}, \citenamefont {Xu}, \citenamefont {Ning},\ and\ \citenamefont
  {Wang}}]{qin2023one}%
  \BibitemOpen
  \bibfield  {author} {\bibinfo {author} {\bibfnamefont {Z.}~\bibnamefont
  {Qin}}, \bibinfo {author} {\bibfnamefont {D.-H.}\ \bibnamefont {Xu}},
  \bibinfo {author} {\bibfnamefont {Z.}~\bibnamefont {Ning}},\ and\ \bibinfo
  {author} {\bibfnamefont {R.}~\bibnamefont {Wang}},\ }\href@noop {} {\bibfield
   {journal} {\bibinfo  {journal} {Physical Review B}\ }\textbf {\bibinfo
  {volume} {108}},\ \bibinfo {pages} {195103} (\bibinfo {year}
  {2023}{\natexlab{b}})}\BibitemShut {NoStop}%
\bibitem [{\citenamefont {Meier}\ \emph {et~al.}(2016)\citenamefont {Meier},
  \citenamefont {An},\ and\ \citenamefont {Gadway}}]{meier2016observation}%
  \BibitemOpen
  \bibfield  {author} {\bibinfo {author} {\bibfnamefont {E.~J.}\ \bibnamefont
  {Meier}}, \bibinfo {author} {\bibfnamefont {F.~A.}\ \bibnamefont {An}},\ and\
  \bibinfo {author} {\bibfnamefont {B.}~\bibnamefont {Gadway}},\ }\href@noop {}
  {\bibfield  {journal} {\bibinfo  {journal} {Nature communications}\ }\textbf
  {\bibinfo {volume} {7}},\ \bibinfo {pages} {13986} (\bibinfo {year}
  {2016})}\BibitemShut {NoStop}%
\bibitem [{ZUM(1984)}]{ZUMINO1984477}%
  \BibitemOpen
  \href {https://doi.org/https://doi.org/10.1016/0550-3213(84)90259-1}
  {\bibfield  {journal} {\bibinfo  {journal} {Nuclear Physics B}\ }\textbf
  {\bibinfo {volume} {239}},\ \bibinfo {pages} {477} (\bibinfo {year}
  {1984})}\BibitemShut {NoStop}%
\bibitem [{\citenamefont {Thouless}(1983)}]{PhysRevB.27.6083}%
  \BibitemOpen
  \bibfield  {author} {\bibinfo {author} {\bibfnamefont {D.~J.}\ \bibnamefont
  {Thouless}},\ }\href {https://doi.org/10.1103/PhysRevB.27.6083} {\bibfield
  {journal} {\bibinfo  {journal} {Phys. Rev. B}\ }\textbf {\bibinfo {volume}
  {27}},\ \bibinfo {pages} {6083} (\bibinfo {year} {1983})}\BibitemShut
  {NoStop}%
\bibitem [{\citenamefont {Niu}(1990)}]{PhysRevLett.64.1812}%
  \BibitemOpen
  \bibfield  {author} {\bibinfo {author} {\bibfnamefont {Q.}~\bibnamefont
  {Niu}},\ }\href {https://doi.org/10.1103/PhysRevLett.64.1812} {\bibfield
  {journal} {\bibinfo  {journal} {Phys. Rev. Lett.}\ }\textbf {\bibinfo
  {volume} {64}},\ \bibinfo {pages} {1812} (\bibinfo {year}
  {1990})}\BibitemShut {NoStop}%
\bibitem [{\citenamefont {Lohse}\ \emph {et~al.}(2016)\citenamefont {Lohse},
  \citenamefont {Schweizer}, \citenamefont {Zilberberg}, \citenamefont
  {Aidelsburger},\ and\ \citenamefont {Bloch}}]{Lohse2016}%
  \BibitemOpen
  \bibfield  {author} {\bibinfo {author} {\bibfnamefont {M.}~\bibnamefont
  {Lohse}}, \bibinfo {author} {\bibfnamefont {C.}~\bibnamefont {Schweizer}},
  \bibinfo {author} {\bibfnamefont {O.}~\bibnamefont {Zilberberg}}, \bibinfo
  {author} {\bibfnamefont {M.}~\bibnamefont {Aidelsburger}},\ and\ \bibinfo
  {author} {\bibfnamefont {I.}~\bibnamefont {Bloch}},\ }\href
  {https://doi.org/10.1038/nphys3584} {\bibfield  {journal} {\bibinfo
  {journal} {Nature Physics}\ }\textbf {\bibinfo {volume} {12}},\ \bibinfo
  {pages} {350} (\bibinfo {year} {2016})}\BibitemShut {NoStop}%
\bibitem [{\citenamefont {Nakajima}\ \emph {et~al.}(2016)\citenamefont
  {Nakajima}, \citenamefont {Tomita}, \citenamefont {Taie}, \citenamefont
  {Ichinose}, \citenamefont {Ozawa}, \citenamefont {Wang}, \citenamefont
  {Troyer},\ and\ \citenamefont {Takahashi}}]{Nakajima2016}%
  \BibitemOpen
  \bibfield  {author} {\bibinfo {author} {\bibfnamefont {S.}~\bibnamefont
  {Nakajima}}, \bibinfo {author} {\bibfnamefont {T.}~\bibnamefont {Tomita}},
  \bibinfo {author} {\bibfnamefont {S.}~\bibnamefont {Taie}}, \bibinfo {author}
  {\bibfnamefont {T.}~\bibnamefont {Ichinose}}, \bibinfo {author}
  {\bibfnamefont {H.}~\bibnamefont {Ozawa}}, \bibinfo {author} {\bibfnamefont
  {L.}~\bibnamefont {Wang}}, \bibinfo {author} {\bibfnamefont {M.}~\bibnamefont
  {Troyer}},\ and\ \bibinfo {author} {\bibfnamefont {Y.}~\bibnamefont
  {Takahashi}},\ }\href {https://doi.org/10.1038/nphys3622} {\bibfield
  {journal} {\bibinfo  {journal} {Nature Physics}\ }\textbf {\bibinfo {volume}
  {12}},\ \bibinfo {pages} {296} (\bibinfo {year} {2016})}\BibitemShut
  {NoStop}%
\bibitem [{\citenamefont {Bevan}\ \emph {et~al.}(1997)\citenamefont {Bevan},
  \citenamefont {Manninen}, \citenamefont {Cook}, \citenamefont {Hook},
  \citenamefont {Hall}, \citenamefont {Vachaspati},\ and\ \citenamefont
  {Volovik}}]{bevan1997momentum}%
  \BibitemOpen
  \bibfield  {author} {\bibinfo {author} {\bibfnamefont {T.}~\bibnamefont
  {Bevan}}, \bibinfo {author} {\bibfnamefont {A.}~\bibnamefont {Manninen}},
  \bibinfo {author} {\bibfnamefont {J.}~\bibnamefont {Cook}}, \bibinfo {author}
  {\bibfnamefont {J.}~\bibnamefont {Hook}}, \bibinfo {author} {\bibfnamefont
  {H.}~\bibnamefont {Hall}}, \bibinfo {author} {\bibfnamefont {T.}~\bibnamefont
  {Vachaspati}},\ and\ \bibinfo {author} {\bibfnamefont {G.}~\bibnamefont
  {Volovik}},\ }\href@noop {} {\bibfield  {journal} {\bibinfo  {journal}
  {Nature}\ }\textbf {\bibinfo {volume} {386}},\ \bibinfo {pages} {689}
  (\bibinfo {year} {1997})}\BibitemShut {NoStop}%
\bibitem [{\citenamefont {Gusynin}\ \emph {et~al.}(1999)\citenamefont
  {Gusynin}, \citenamefont {Miransky},\ and\ \citenamefont
  {Shovkovy}}]{Gusynin1999pq}%
  \BibitemOpen
  \bibfield  {author} {\bibinfo {author} {\bibfnamefont {V.~P.}\ \bibnamefont
  {Gusynin}}, \bibinfo {author} {\bibfnamefont {V.~A.}\ \bibnamefont
  {Miransky}},\ and\ \bibinfo {author} {\bibfnamefont {I.~A.}\ \bibnamefont
  {Shovkovy}},\ }\href {https://doi.org/10.1016/S0550-3213(99)00573-8}
  {\bibfield  {journal} {\bibinfo  {journal} {Nucl. Phys. B}\ }\textbf
  {\bibinfo {volume} {563}},\ \bibinfo {pages} {361} (\bibinfo {year}
  {1999})},\ \Eprint {https://arxiv.org/abs/hep-ph/9908320}
  {arXiv:hep-ph/9908320} \BibitemShut {NoStop}%
\bibitem [{\citenamefont {Abramchuk}\ and\ \citenamefont
  {Zubkov}(2024)}]{abramchuk2024magnetoconductivity}%
  \BibitemOpen
  \bibfield  {author} {\bibinfo {author} {\bibfnamefont {R.}~\bibnamefont
  {Abramchuk}}\ and\ \bibinfo {author} {\bibfnamefont {M.}~\bibnamefont
  {Zubkov}},\ }\href@noop {} {\bibinfo {title} {Magnetoconductivity of dirac
  semimetals and chiral magnetic effect from keldysh technique}} (\bibinfo
  {year} {2024}),\ \bibinfo {note} {arXiv preprint
  arXiv:2409.14941}\BibitemShut {NoStop}%
\bibitem [{\citenamefont {Abramchuk}(2026)}]{ABRAMCHUK2026113374}%
  \BibitemOpen
  \bibfield  {author} {\bibinfo {author} {\bibfnamefont {R.~A.}\ \bibnamefont
  {Abramchuk}},\ }\href
  {https://doi.org/https://doi.org/10.1016/j.jpcs.2025.113374} {\bibfield
  {journal} {\bibinfo  {journal} {Journal of Physics and Chemistry of Solids}\
  }\textbf {\bibinfo {volume} {210}},\ \bibinfo {pages} {113374} (\bibinfo
  {year} {2026})}\BibitemShut {NoStop}%
\bibitem [{\citenamefont {Li}\ \emph {et~al.}(2016)\citenamefont {Li},
  \citenamefont {Kharzeev}, \citenamefont {Zhang}, \citenamefont {Huang},
  \citenamefont {Pletikosic}, \citenamefont {Fedorov}, \citenamefont {Zhong},
  \citenamefont {Schneeloch}, \citenamefont {Gu},\ and\ \citenamefont
  {Valla}}]{CMEZrTe5}%
  \BibitemOpen
  \bibfield  {author} {\bibinfo {author} {\bibfnamefont {Q.}~\bibnamefont
  {Li}}, \bibinfo {author} {\bibfnamefont {D.~E.}\ \bibnamefont {Kharzeev}},
  \bibinfo {author} {\bibfnamefont {C.}~\bibnamefont {Zhang}}, \bibinfo
  {author} {\bibfnamefont {Y.}~\bibnamefont {Huang}}, \bibinfo {author}
  {\bibfnamefont {I.}~\bibnamefont {Pletikosic}}, \bibinfo {author}
  {\bibfnamefont {A.~V.}\ \bibnamefont {Fedorov}}, \bibinfo {author}
  {\bibfnamefont {R.~D.}\ \bibnamefont {Zhong}}, \bibinfo {author}
  {\bibfnamefont {J.~A.}\ \bibnamefont {Schneeloch}}, \bibinfo {author}
  {\bibfnamefont {G.~D.}\ \bibnamefont {Gu}},\ and\ \bibinfo {author}
  {\bibfnamefont {T.}~\bibnamefont {Valla}},\ }\href
  {https://doi.org/10.1038/nphys3648} {\bibfield  {journal} {\bibinfo
  {journal} {Nature Phys.}\ }\textbf {\bibinfo {volume} {12}},\ \bibinfo
  {pages} {550} (\bibinfo {year} {2016})},\ \Eprint
  {https://arxiv.org/abs/1412.6543} {arXiv:1412.6543 [cond-mat.str-el]}
  \BibitemShut {NoStop}%
\bibitem [{\citenamefont {Kaushik}\ and\ \citenamefont
  {Kharzeev}(2017)}]{Kharzeev2017}%
  \BibitemOpen
  \bibfield  {author} {\bibinfo {author} {\bibfnamefont {S.}~\bibnamefont
  {Kaushik}}\ and\ \bibinfo {author} {\bibfnamefont {D.~E.}\ \bibnamefont
  {Kharzeev}},\ }\href {https://doi.org/10.1103/PhysRevB.95.235136} {\bibfield
  {journal} {\bibinfo  {journal} {Phys. Rev. B}\ }\textbf {\bibinfo {volume}
  {95}},\ \bibinfo {pages} {235136} (\bibinfo {year} {2017})},\ \Eprint
  {https://arxiv.org/abs/1703.05865} {arXiv:1703.05865 [cond-mat.mes-hall]}
  \BibitemShut {NoStop}%
\bibitem [{\citenamefont {Chernodub}\ and\ \citenamefont
  {Zubkov}(2017)}]{chernodub2017scale}%
  \BibitemOpen
  \bibfield  {author} {\bibinfo {author} {\bibfnamefont {M.~N.}\ \bibnamefont
  {Chernodub}}\ and\ \bibinfo {author} {\bibfnamefont {M.}~\bibnamefont
  {Zubkov}},\ }\href@noop {} {\bibfield  {journal} {\bibinfo  {journal}
  {Physical Review D}\ }\textbf {\bibinfo {volume} {96}},\ \bibinfo {pages}
  {056006} (\bibinfo {year} {2017})}\BibitemShut {NoStop}%
\bibitem [{\citenamefont {Zhang}\ and\ \citenamefont
  {Zubkov}(2020)}]{zhang2020influence}%
  \BibitemOpen
  \bibfield  {author} {\bibinfo {author} {\bibfnamefont {C.}~\bibnamefont
  {Zhang}}\ and\ \bibinfo {author} {\bibfnamefont {M.}~\bibnamefont {Zubkov}},\
  }\href@noop {} {\bibfield  {journal} {\bibinfo  {journal} {Journal of Physics
  A: Mathematical and Theoretical}\ }\textbf {\bibinfo {volume} {53}},\
  \bibinfo {pages} {195002} (\bibinfo {year} {2020})}\BibitemShut {NoStop}%
\bibitem [{\citenamefont {Zubkov}\ and\ \citenamefont
  {Abramchuk}(2023)}]{zubkov2023effect}%
  \BibitemOpen
  \bibfield  {author} {\bibinfo {author} {\bibfnamefont {M.}~\bibnamefont
  {Zubkov}}\ and\ \bibinfo {author} {\bibfnamefont {R.}~\bibnamefont
  {Abramchuk}},\ }\href@noop {} {\bibfield  {journal} {\bibinfo  {journal}
  {Physical Review D}\ }\textbf {\bibinfo {volume} {107}},\ \bibinfo {pages}
  {094021} (\bibinfo {year} {2023})}\BibitemShut {NoStop}%
\bibitem [{\citenamefont {Abramchuk}\ \emph {et~al.}(2018)\citenamefont
  {Abramchuk}, \citenamefont {Khaidukov},\ and\ \citenamefont
  {Zubkov}}]{abramchuk2018anatomy}%
  \BibitemOpen
  \bibfield  {author} {\bibinfo {author} {\bibfnamefont {R.}~\bibnamefont
  {Abramchuk}}, \bibinfo {author} {\bibfnamefont {Z.}~\bibnamefont
  {Khaidukov}},\ and\ \bibinfo {author} {\bibfnamefont {M.}~\bibnamefont
  {Zubkov}},\ }\href@noop {} {\bibfield  {journal} {\bibinfo  {journal}
  {Physical Review D}\ }\textbf {\bibinfo {volume} {98}},\ \bibinfo {pages}
  {076013} (\bibinfo {year} {2018})}\BibitemShut {NoStop}%
\bibitem [{\citenamefont {Volovik}(2003)}]{Volovik2003a}%
  \BibitemOpen
  \bibfield  {author} {\bibinfo {author} {\bibfnamefont {G.~E.}\ \bibnamefont
  {Volovik}},\ }\href@noop {} {\emph {\bibinfo {title} {{The Universe in a
  Helium Droplet}}}}\ (\bibinfo  {publisher} {Clarendon Press},\ \bibinfo
  {address} {Oxford},\ \bibinfo {year} {2003})\BibitemShut {NoStop}%
\bibitem [{\citenamefont {Zubkov}(2018)}]{zubkov2018momentum}%
  \BibitemOpen
  \bibfield  {author} {\bibinfo {author} {\bibfnamefont {M.}~\bibnamefont
  {Zubkov}},\ }\href@noop {} {\bibfield  {journal} {\bibinfo  {journal} {Annals
  of Physics}\ }\textbf {\bibinfo {volume} {393}},\ \bibinfo {pages} {264}
  (\bibinfo {year} {2018})}\BibitemShut {NoStop}%
\bibitem [{\citenamefont {Zubkov}\ and\ \citenamefont
  {Volovik}(2012)}]{zubkov2012momentum}%
  \BibitemOpen
  \bibfield  {author} {\bibinfo {author} {\bibfnamefont {M.}~\bibnamefont
  {Zubkov}}\ and\ \bibinfo {author} {\bibfnamefont {G.}~\bibnamefont
  {Volovik}},\ }\href@noop {} {\bibfield  {journal} {\bibinfo  {journal}
  {Nuclear Physics B}\ }\textbf {\bibinfo {volume} {860}},\ \bibinfo {pages}
  {295} (\bibinfo {year} {2012})}\BibitemShut {NoStop}%
\bibitem [{\citenamefont {Volovik}\ and\ \citenamefont
  {Zubkov}(2017)}]{volovik2017standard}%
  \BibitemOpen
  \bibfield  {author} {\bibinfo {author} {\bibfnamefont {G.}~\bibnamefont
  {Volovik}}\ and\ \bibinfo {author} {\bibfnamefont {M.}~\bibnamefont
  {Zubkov}},\ }\href@noop {} {\bibfield  {journal} {\bibinfo  {journal} {New
  Journal of Physics}\ }\textbf {\bibinfo {volume} {19}},\ \bibinfo {pages}
  {015009} (\bibinfo {year} {2017})}\BibitemShut {NoStop}%
\bibitem [{\citenamefont {Volovik}\ and\ \citenamefont
  {Zubkov}(2013)}]{volovik2013nambu}%
  \BibitemOpen
  \bibfield  {author} {\bibinfo {author} {\bibfnamefont {G.~E.}\ \bibnamefont
  {Volovik}}\ and\ \bibinfo {author} {\bibfnamefont {M.}~\bibnamefont
  {Zubkov}},\ }\href@noop {} {\bibfield  {journal} {\bibinfo  {journal} {JETP
  letters}\ }\textbf {\bibinfo {volume} {97}},\ \bibinfo {pages} {301}
  (\bibinfo {year} {2013})}\BibitemShut {NoStop}%
\bibitem [{\citenamefont {Volovik}\ and\ \citenamefont
  {Zubkov}(2015)}]{volovik2015scalar}%
  \BibitemOpen
  \bibfield  {author} {\bibinfo {author} {\bibfnamefont {G.}~\bibnamefont
  {Volovik}}\ and\ \bibinfo {author} {\bibfnamefont {M.}~\bibnamefont
  {Zubkov}},\ }\href@noop {} {\bibfield  {journal} {\bibinfo  {journal}
  {Physical Review D}\ }\textbf {\bibinfo {volume} {92}},\ \bibinfo {pages}
  {055004} (\bibinfo {year} {2015})}\BibitemShut {NoStop}%
\bibitem [{\citenamefont {Z.Khaidukov}\ and\ \citenamefont
  {M.A.Zubkov}(2017)}]{zubkov2017topology}%
  \BibitemOpen
  \bibfield  {author} {\bibinfo {author} {\bibnamefont {Z.Khaidukov}}\ and\
  \bibinfo {author} {\bibnamefont {M.A.Zubkov}},\ }\href@noop {} {\bibfield
  {journal} {\bibinfo  {journal} {JETP Letters}\ }\textbf {\bibinfo {volume}
  {106}},\ \bibinfo {pages} {172} (\bibinfo {year} {2017})}\BibitemShut
  {NoStop}%
\bibitem [{\citenamefont {Zhang}\ and\ \citenamefont
  {Zubkov}(2019)}]{zhang2019hall}%
  \BibitemOpen
  \bibfield  {author} {\bibinfo {author} {\bibfnamefont {C.}~\bibnamefont
  {Zhang}}\ and\ \bibinfo {author} {\bibfnamefont {M.}~\bibnamefont {Zubkov}},\
  }\href@noop {} {\bibfield  {journal} {\bibinfo  {journal} {JETP letters}\
  }\textbf {\bibinfo {volume} {110}},\ \bibinfo {pages} {487} (\bibinfo {year}
  {2019})}\BibitemShut {NoStop}%
\bibitem [{\citenamefont {Selch}\ \emph {et~al.}(2025)\citenamefont {Selch},
  \citenamefont {Zubkov}, \citenamefont {Pramanik},\ and\ \citenamefont
  {Lewkowicz}}]{selch2025non}%
  \BibitemOpen
  \bibfield  {author} {\bibinfo {author} {\bibfnamefont {M.}~\bibnamefont
  {Selch}}, \bibinfo {author} {\bibfnamefont {M.}~\bibnamefont {Zubkov}},
  \bibinfo {author} {\bibfnamefont {S.}~\bibnamefont {Pramanik}},\ and\
  \bibinfo {author} {\bibfnamefont {M.}~\bibnamefont {Lewkowicz}},\ }\href@noop
  {} {\bibfield  {journal} {\bibinfo  {journal} {Annals of Physics}\ ,\
  \bibinfo {pages} {170202}} (\bibinfo {year} {2025})}\BibitemShut {NoStop}%
\bibitem [{\citenamefont {Katsnelson}\ \emph {et~al.}(2013)\citenamefont
  {Katsnelson}, \citenamefont {Volovik},\ and\ \citenamefont
  {Zubkov}}]{katsnelson2013euler}%
  \BibitemOpen
  \bibfield  {author} {\bibinfo {author} {\bibfnamefont {M.}~\bibnamefont
  {Katsnelson}}, \bibinfo {author} {\bibfnamefont {G.}~\bibnamefont
  {Volovik}},\ and\ \bibinfo {author} {\bibfnamefont {M.}~\bibnamefont
  {Zubkov}},\ }\href@noop {} {\bibfield  {journal} {\bibinfo  {journal} {Annals
  of Physics}\ }\textbf {\bibinfo {volume} {331}},\ \bibinfo {pages} {160}
  (\bibinfo {year} {2013})}\BibitemShut {NoStop}%
\bibitem [{\citenamefont {Bakker}\ \emph {et~al.}(1999)\citenamefont {Bakker},
  \citenamefont {Veselov},\ and\ \citenamefont {Zubkov}}]{bakker1999central}%
  \BibitemOpen
  \bibfield  {author} {\bibinfo {author} {\bibfnamefont {B.}~\bibnamefont
  {Bakker}}, \bibinfo {author} {\bibfnamefont {A.}~\bibnamefont {Veselov}},\
  and\ \bibinfo {author} {\bibfnamefont {M.}~\bibnamefont {Zubkov}},\
  }\href@noop {} {\bibfield  {journal} {\bibinfo  {journal} {Physics Letters
  B}\ }\textbf {\bibinfo {volume} {471}},\ \bibinfo {pages} {214} (\bibinfo
  {year} {1999})}\BibitemShut {NoStop}%
\bibitem [{\citenamefont {Bakker}\ \emph {et~al.}(2005)\citenamefont {Bakker},
  \citenamefont {Veselov},\ and\ \citenamefont {Zubkov}}]{bakker2005standard}%
  \BibitemOpen
  \bibfield  {author} {\bibinfo {author} {\bibfnamefont {B.}~\bibnamefont
  {Bakker}}, \bibinfo {author} {\bibfnamefont {A.}~\bibnamefont {Veselov}},\
  and\ \bibinfo {author} {\bibfnamefont {M.}~\bibnamefont {Zubkov}},\
  }\href@noop {} {\bibfield  {journal} {\bibinfo  {journal} {Physics Letters
  B}\ }\textbf {\bibinfo {volume} {620}},\ \bibinfo {pages} {156} (\bibinfo
  {year} {2005})}\BibitemShut {NoStop}%
\bibitem [{\citenamefont {Bohra}\ and\ \citenamefont
  {Zubkov}(2025)}]{bohra2025relation}%
  \BibitemOpen
  \bibfield  {author} {\bibinfo {author} {\bibfnamefont {M.}~\bibnamefont
  {Bohra}}\ and\ \bibinfo {author} {\bibfnamefont {M.}~\bibnamefont {Zubkov}},\
  }\href@noop {} {\bibfield  {journal} {\bibinfo  {journal} {Solid State
  Communications}\ ,\ \bibinfo {pages} {116276}} (\bibinfo {year}
  {2025})}\BibitemShut {NoStop}%
\bibitem [{\citenamefont {Fukushima}\ \emph {et~al.}(2008)\citenamefont
  {Fukushima}, \citenamefont {Kharzeev},\ and\ \citenamefont
  {Warringa}}]{Fukushima2008}%
  \BibitemOpen
  \bibfield  {author} {\bibinfo {author} {\bibfnamefont {K.}~\bibnamefont
  {Fukushima}}, \bibinfo {author} {\bibfnamefont {D.~E.}\ \bibnamefont
  {Kharzeev}},\ and\ \bibinfo {author} {\bibfnamefont {H.~J.}\ \bibnamefont
  {Warringa}},\ }\href {https://doi.org/10.1103/PhysRevD.78.074033} {\bibfield
  {journal} {\bibinfo  {journal} {Phys. Rev. D}\ }\textbf {\bibinfo {volume}
  {78}},\ \bibinfo {pages} {074033} (\bibinfo {year} {2008})},\ \Eprint
  {https://arxiv.org/abs/0808.3382} {arXiv:0808.3382 [hep-ph]} \BibitemShut
  {NoStop}%
\bibitem [{\citenamefont {Stephanov}\ and\ \citenamefont
  {Yin}(2012)}]{Stephanov2012ki}%
  \BibitemOpen
  \bibfield  {author} {\bibinfo {author} {\bibfnamefont {M.~A.}\ \bibnamefont
  {Stephanov}}\ and\ \bibinfo {author} {\bibfnamefont {Y.}~\bibnamefont
  {Yin}},\ }\href {https://doi.org/10.1103/PhysRevLett.109.162001} {\bibfield
  {journal} {\bibinfo  {journal} {Phys. Rev. Lett.}\ }\textbf {\bibinfo
  {volume} {109}},\ \bibinfo {pages} {162001} (\bibinfo {year} {2012})},\
  \Eprint {https://arxiv.org/abs/1207.0747} {arXiv:1207.0747 [hep-th]}
  \BibitemShut {NoStop}%
\bibitem [{\citenamefont {Burkov}(2014)}]{Burkov2014}%
  \BibitemOpen
  \bibfield  {author} {\bibinfo {author} {\bibfnamefont {A.}~\bibnamefont
  {Burkov}},\ }\bibfield  {journal} {\bibinfo  {journal} {Physical Review
  Letters}\ }\textbf {\bibinfo {volume} {113}},\ \href
  {https://doi.org/10.1103/physrevlett.113.247203}
  {10.1103/physrevlett.113.247203} (\bibinfo {year} {2014}),\ \Eprint
  {https://arxiv.org/abs/1409.0013} {arXiv:1409.0013} \BibitemShut {NoStop}%
\bibitem [{\citenamefont {Son}\ and\ \citenamefont {Spivak}(2013)}]{Son2013}%
  \BibitemOpen
  \bibfield  {author} {\bibinfo {author} {\bibfnamefont {D.~T.}\ \bibnamefont
  {Son}}\ and\ \bibinfo {author} {\bibfnamefont {B.~Z.}\ \bibnamefont
  {Spivak}},\ }\bibfield  {journal} {\bibinfo  {journal} {Physical Review B}\
  }\textbf {\bibinfo {volume} {88}},\ \href
  {https://doi.org/10.1103/physrevb.88.104412} {10.1103/physrevb.88.104412}
  (\bibinfo {year} {2013}),\ \Eprint {https://arxiv.org/abs/1206.1627}
  {arXiv:1206.1627} \BibitemShut {NoStop}%
\bibitem [{\citenamefont {Gorbar}\ \emph {et~al.}(2016)\citenamefont {Gorbar},
  \citenamefont {Shovkovy}, \citenamefont {Vilchinskii}, \citenamefont
  {Rudenok}, \citenamefont {Boyarsky},\ and\ \citenamefont
  {Ruchayskiy}}]{Gorbar2016}%
  \BibitemOpen
  \bibfield  {author} {\bibinfo {author} {\bibfnamefont {E.~V.}\ \bibnamefont
  {Gorbar}}, \bibinfo {author} {\bibfnamefont {I.~A.}\ \bibnamefont
  {Shovkovy}}, \bibinfo {author} {\bibfnamefont {S.}~\bibnamefont
  {Vilchinskii}}, \bibinfo {author} {\bibfnamefont {I.}~\bibnamefont
  {Rudenok}}, \bibinfo {author} {\bibfnamefont {A.}~\bibnamefont {Boyarsky}},\
  and\ \bibinfo {author} {\bibfnamefont {O.}~\bibnamefont {Ruchayskiy}},\
  }\href {https://doi.org/10.1103/PhysRevD.93.105028} {\bibfield  {journal}
  {\bibinfo  {journal} {Phys. Rev. D}\ }\textbf {\bibinfo {volume} {93}},\
  \bibinfo {pages} {105028} (\bibinfo {year} {2016})},\ \Eprint
  {https://arxiv.org/abs/1603.03442} {arXiv:1603.03442 [hep-th]} \BibitemShut
  {NoStop}%
\bibitem [{\citenamefont {Gao}\ \emph {et~al.}(2012)\citenamefont {Gao},
  \citenamefont {Liang}, \citenamefont {Pu}, \citenamefont {Wang},\ and\
  \citenamefont {Wang}}]{Gao2012}%
  \BibitemOpen
  \bibfield  {author} {\bibinfo {author} {\bibfnamefont {J.-H.}\ \bibnamefont
  {Gao}}, \bibinfo {author} {\bibfnamefont {Z.-T.}\ \bibnamefont {Liang}},
  \bibinfo {author} {\bibfnamefont {S.}~\bibnamefont {Pu}}, \bibinfo {author}
  {\bibfnamefont {Q.}~\bibnamefont {Wang}},\ and\ \bibinfo {author}
  {\bibfnamefont {X.-N.}\ \bibnamefont {Wang}},\ }\href
  {https://doi.org/10.1103/PhysRevLett.109.232301} {\bibfield  {journal}
  {\bibinfo  {journal} {Phys. Rev. Lett.}\ }\textbf {\bibinfo {volume} {109}},\
  \bibinfo {pages} {232301} (\bibinfo {year} {2012})},\ \Eprint
  {https://arxiv.org/abs/1203.0725} {arXiv:1203.0725 [hep-ph]} \BibitemShut
  {NoStop}%
\bibitem [{\citenamefont {Lin}\ and\ \citenamefont {Yang}(2020)}]{Lin2019}%
  \BibitemOpen
  \bibfield  {author} {\bibinfo {author} {\bibfnamefont {S.}~\bibnamefont
  {Lin}}\ and\ \bibinfo {author} {\bibfnamefont {L.}~\bibnamefont {Yang}},\
  }\href {https://doi.org/10.1103/PhysRevD.101.034006} {\bibfield  {journal}
  {\bibinfo  {journal} {Phys. Rev. D}\ }\textbf {\bibinfo {volume} {101}},\
  \bibinfo {pages} {034006} (\bibinfo {year} {2020})},\ \Eprint
  {https://arxiv.org/abs/1909.11514} {arXiv:1909.11514 [nucl-th]} \BibitemShut
  {NoStop}%
\bibitem [{\citenamefont {Hattori}\ \emph {et~al.}(2017)\citenamefont
  {Hattori}, \citenamefont {Li}, \citenamefont {Satow},\ and\ \citenamefont
  {Yee}}]{Hattori2016lqx}%
  \BibitemOpen
  \bibfield  {author} {\bibinfo {author} {\bibfnamefont {K.}~\bibnamefont
  {Hattori}}, \bibinfo {author} {\bibfnamefont {S.}~\bibnamefont {Li}},
  \bibinfo {author} {\bibfnamefont {D.}~\bibnamefont {Satow}},\ and\ \bibinfo
  {author} {\bibfnamefont {H.-U.}\ \bibnamefont {Yee}},\ }\href
  {https://doi.org/10.1103/PhysRevD.95.076008} {\bibfield  {journal} {\bibinfo
  {journal} {Phys. Rev. D}\ }\textbf {\bibinfo {volume} {95}},\ \bibinfo
  {pages} {076008} (\bibinfo {year} {2017})},\ \Eprint
  {https://arxiv.org/abs/1610.06839} {arXiv:1610.06839 [hep-ph]} \BibitemShut
  {NoStop}%
\bibitem [{\citenamefont {Huang}\ \emph {et~al.}(2018)\citenamefont {Huang},
  \citenamefont {Jiang}, \citenamefont {Shi}, \citenamefont {Liao},\ and\
  \citenamefont {Zhuang}}]{Huang2017}%
  \BibitemOpen
  \bibfield  {author} {\bibinfo {author} {\bibfnamefont {A.}~\bibnamefont
  {Huang}}, \bibinfo {author} {\bibfnamefont {Y.}~\bibnamefont {Jiang}},
  \bibinfo {author} {\bibfnamefont {S.}~\bibnamefont {Shi}}, \bibinfo {author}
  {\bibfnamefont {J.}~\bibnamefont {Liao}},\ and\ \bibinfo {author}
  {\bibfnamefont {P.}~\bibnamefont {Zhuang}},\ }\href
  {https://doi.org/10.1016/j.physletb.2017.12.025} {\bibfield  {journal}
  {\bibinfo  {journal} {Phys. Lett. B}\ }\textbf {\bibinfo {volume} {777}},\
  \bibinfo {pages} {177} (\bibinfo {year} {2018})},\ \Eprint
  {https://arxiv.org/abs/1703.08856} {arXiv:1703.08856 [hep-ph]} \BibitemShut
  {NoStop}%
\bibitem [{\citenamefont {Son}\ and\ \citenamefont
  {Yamamoto}(2012)}]{Son_2012}%
  \BibitemOpen
  \bibfield  {author} {\bibinfo {author} {\bibfnamefont {D.~T.}\ \bibnamefont
  {Son}}\ and\ \bibinfo {author} {\bibfnamefont {N.}~\bibnamefont {Yamamoto}},\
  }\bibfield  {journal} {\bibinfo  {journal} {Physical Review Letters}\
  }\textbf {\bibinfo {volume} {109}},\ \href
  {https://doi.org/10.1103/physrevlett.109.181602}
  {10.1103/physrevlett.109.181602} (\bibinfo {year} {2012})\BibitemShut
  {NoStop}%
\bibitem [{\citenamefont {Sekine}\ \emph {et~al.}(2017)\citenamefont {Sekine},
  \citenamefont {Culcer},\ and\ \citenamefont
  {MacDonald}}]{PhysRevB.96.235134}%
  \BibitemOpen
  \bibfield  {author} {\bibinfo {author} {\bibfnamefont {A.}~\bibnamefont
  {Sekine}}, \bibinfo {author} {\bibfnamefont {D.}~\bibnamefont {Culcer}},\
  and\ \bibinfo {author} {\bibfnamefont {A.~H.}\ \bibnamefont {MacDonald}},\
  }\href {https://doi.org/10.1103/PhysRevB.96.235134} {\bibfield  {journal}
  {\bibinfo  {journal} {Phys. Rev. B}\ }\textbf {\bibinfo {volume} {96}},\
  \bibinfo {pages} {235134} (\bibinfo {year} {2017})}\BibitemShut {NoStop}%
\bibitem [{\citenamefont {Sekine}\ and\ \citenamefont
  {Nomura}(2021)}]{sekine2021axion}%
  \BibitemOpen
  \bibfield  {author} {\bibinfo {author} {\bibfnamefont {A.}~\bibnamefont
  {Sekine}}\ and\ \bibinfo {author} {\bibfnamefont {K.}~\bibnamefont
  {Nomura}},\ }\href@noop {} {\bibfield  {journal} {\bibinfo  {journal}
  {Journal of Applied Physics}\ }\textbf {\bibinfo {volume} {129}} (\bibinfo
  {year} {2021})}\BibitemShut {NoStop}%
\bibitem [{\citenamefont {Volovik}(2017)}]{volovik2017chiral}%
  \BibitemOpen
  \bibfield  {author} {\bibinfo {author} {\bibfnamefont {G.~E.}\ \bibnamefont
  {Volovik}},\ }\href@noop {} {\bibfield  {journal} {\bibinfo  {journal} {JETP
  letters}\ }\textbf {\bibinfo {volume} {105}},\ \bibinfo {pages} {34}
  (\bibinfo {year} {2017})}\BibitemShut {NoStop}%
\bibitem [{\citenamefont {Suleymanov}\ and\ \citenamefont
  {Zubkov}(2019)}]{suleymanov2019wigner}%
  \BibitemOpen
  \bibfield  {author} {\bibinfo {author} {\bibfnamefont {M.}~\bibnamefont
  {Suleymanov}}\ and\ \bibinfo {author} {\bibfnamefont {M.}~\bibnamefont
  {Zubkov}},\ }\href@noop {} {\bibfield  {journal} {\bibinfo  {journal}
  {Nuclear Physics B}\ }\textbf {\bibinfo {volume} {938}},\ \bibinfo {pages}
  {171} (\bibinfo {year} {2019})}\BibitemShut {NoStop}%
\bibitem [{\citenamefont {Gorbar}\ \emph {et~al.}(2014)\citenamefont {Gorbar},
  \citenamefont {Miransky},\ and\ \citenamefont {Shovkovy}}]{gorbar2014chiral}%
  \BibitemOpen
  \bibfield  {author} {\bibinfo {author} {\bibfnamefont {E.}~\bibnamefont
  {Gorbar}}, \bibinfo {author} {\bibfnamefont {V.}~\bibnamefont {Miransky}},\
  and\ \bibinfo {author} {\bibfnamefont {I.}~\bibnamefont {Shovkovy}},\
  }\href@noop {} {\bibfield  {journal} {\bibinfo  {journal} {Physical Review
  B}\ }\textbf {\bibinfo {volume} {89}},\ \bibinfo {pages} {085126} (\bibinfo
  {year} {2014})},\ \Eprint {https://arxiv.org/abs/1312.0027} {arXiv:1312.0027
  [cond-mat.mes-hall]} \BibitemShut {NoStop}%
\bibitem [{\citenamefont {Lu}\ \emph {et~al.}(2015)\citenamefont {Lu},
  \citenamefont {Zhang},\ and\ \citenamefont {Shen}}]{Lu_2015}%
  \BibitemOpen
  \bibfield  {author} {\bibinfo {author} {\bibfnamefont {H.-Z.}\ \bibnamefont
  {Lu}}, \bibinfo {author} {\bibfnamefont {S.-B.}\ \bibnamefont {Zhang}},\ and\
  \bibinfo {author} {\bibfnamefont {S.-Q.}\ \bibnamefont {Shen}},\ }\href
  {https://doi.org/10.1103/PhysRevB.92.045203} {\bibfield  {journal} {\bibinfo
  {journal} {Phys. Rev. B}\ }\textbf {\bibinfo {volume} {92}},\ \bibinfo
  {pages} {045203} (\bibinfo {year} {2015})},\ \Eprint
  {https://arxiv.org/abs/1503.04394} {arXiv:1503.04394} \BibitemShut {NoStop}%
\bibitem [{\citenamefont {Li}\ \emph {et~al.}(2023)\citenamefont {Li},
  \citenamefont {Lu},\ and\ \citenamefont {Xie}}]{Li_2023}%
  \BibitemOpen
  \bibfield  {author} {\bibinfo {author} {\bibfnamefont {S.}~\bibnamefont
  {Li}}, \bibinfo {author} {\bibfnamefont {H.-Z.}\ \bibnamefont {Lu}},\ and\
  \bibinfo {author} {\bibfnamefont {X.~C.}\ \bibnamefont {Xie}},\ }\bibfield
  {journal} {\bibinfo  {journal} {Physical Review B}\ }\textbf {\bibinfo
  {volume} {107}},\ \href {https://doi.org/10.1103/physrevb.107.235202}
  {10.1103/physrevb.107.235202} (\bibinfo {year} {2023}),\ \Eprint
  {https://arxiv.org/abs/2212.00383} {arXiv:2212.00383} \BibitemShut {NoStop}%
\bibitem [{\citenamefont {Zhang}\ and\ \citenamefont
  {Zhou}(2017)}]{PhysRevA.95.061601}%
  \BibitemOpen
  \bibfield  {author} {\bibinfo {author} {\bibfnamefont {S.-L.}\ \bibnamefont
  {Zhang}}\ and\ \bibinfo {author} {\bibfnamefont {Q.}~\bibnamefont {Zhou}},\
  }\href {https://doi.org/10.1103/PhysRevA.95.061601} {\bibfield  {journal}
  {\bibinfo  {journal} {Phys. Rev. A}\ }\textbf {\bibinfo {volume} {95}},\
  \bibinfo {pages} {061601} (\bibinfo {year} {2017})}\BibitemShut {NoStop}%
\bibitem [{\citenamefont {Blankenbecler}\ \emph {et~al.}(1981)\citenamefont
  {Blankenbecler}, \citenamefont {Scalapino},\ and\ \citenamefont
  {Sugar}}]{Blankenbecler81}%
  \BibitemOpen
  \bibfield  {author} {\bibinfo {author} {\bibfnamefont {R.}~\bibnamefont
  {Blankenbecler}}, \bibinfo {author} {\bibfnamefont {D.~J.}\ \bibnamefont
  {Scalapino}},\ and\ \bibinfo {author} {\bibfnamefont {R.~L.}\ \bibnamefont
  {Sugar}},\ }\href {https://doi.org/10.1103/PhysRevD.24.2278} {\bibfield
  {journal} {\bibinfo  {journal} {Phys. Rev. D}\ }\textbf {\bibinfo {volume}
  {24}},\ \bibinfo {pages} {2278} (\bibinfo {year} {1981})}\BibitemShut
  {NoStop}%
\bibitem [{\citenamefont {White}\ \emph {et~al.}(1989)\citenamefont {White},
  \citenamefont {Scalapino}, \citenamefont {Sugar}, \citenamefont {Loh},
  \citenamefont {Gubernatis},\ and\ \citenamefont {Scalettar}}]{White89}%
  \BibitemOpen
  \bibfield  {author} {\bibinfo {author} {\bibfnamefont {S.}~\bibnamefont
  {White}}, \bibinfo {author} {\bibfnamefont {D.}~\bibnamefont {Scalapino}},
  \bibinfo {author} {\bibfnamefont {R.}~\bibnamefont {Sugar}}, \bibinfo
  {author} {\bibfnamefont {E.}~\bibnamefont {Loh}}, \bibinfo {author}
  {\bibfnamefont {J.}~\bibnamefont {Gubernatis}},\ and\ \bibinfo {author}
  {\bibfnamefont {R.}~\bibnamefont {Scalettar}},\ }\href
  {https://doi.org/10.1103/PhysRevB.40.506} {\bibfield  {journal} {\bibinfo
  {journal} {Phys. Rev. B}\ }\textbf {\bibinfo {volume} {40}},\ \bibinfo
  {pages} {506} (\bibinfo {year} {1989})}\BibitemShut {NoStop}%
\bibitem [{\citenamefont {Assaad}\ and\ \citenamefont
  {Evertz}(2008)}]{Assaad08_rev}%
  \BibitemOpen
  \bibfield  {author} {\bibinfo {author} {\bibfnamefont {F.}~\bibnamefont
  {Assaad}}\ and\ \bibinfo {author} {\bibfnamefont {H.}~\bibnamefont
  {Evertz}},\ }in\ \href {https://doi.org/10.1007/978-3-540-74686-7_10} {\emph
  {\bibinfo {booktitle} {Computational Many-Particle Physics}}},\ \bibinfo
  {series} {Lecture Notes in Physics}, Vol.\ \bibinfo {volume} {739},\ \bibinfo
  {editor} {edited by\ \bibinfo {editor} {\bibfnamefont {H.}~\bibnamefont
  {Fehske}}, \bibinfo {editor} {\bibfnamefont {R.}~\bibnamefont {Schneider}},\
  and\ \bibinfo {editor} {\bibfnamefont {A.}~\bibnamefont {Wei{\ss}e}}}\
  (\bibinfo  {publisher} {Springer},\ \bibinfo {address} {Berlin Heidelberg},\
  \bibinfo {year} {2008})\ pp.\ \bibinfo {pages} {277--356}\BibitemShut
  {NoStop}%
\bibitem [{\citenamefont {Ulybyshev}\ \emph {et~al.}(2025)\citenamefont
  {Ulybyshev}, \citenamefont {Reingruber},\ and\ \citenamefont
  {Thremer}}]{superlat_github}%
  \BibitemOpen
  \bibfield  {author} {\bibinfo {author} {\bibfnamefont {M.}~\bibnamefont
  {Ulybyshev}}, \bibinfo {author} {\bibfnamefont {A.}~\bibnamefont
  {Reingruber}},\ and\ \bibinfo {author} {\bibfnamefont {L.}~\bibnamefont
  {Thremer}},\ }\href@noop {} {\bibinfo {title} {Superlattice: Quantum monte
  carlo simulations}},\ \bibinfo {howpublished}
  {\url{https://github.com/ulybyshev/SuperLattice-public}} (\bibinfo {year}
  {2025})\BibitemShut {NoStop}%
\bibitem [{\citenamefont {{Beach}}(2004)}]{Beach04a}%
  \BibitemOpen
  \bibfield  {author} {\bibinfo {author} {\bibfnamefont {K.~S.~D.}\
  \bibnamefont {{Beach}}},\ }\href
  {https://doi.org/10.48550/arXiv.cond-mat/0403055} {\bibfield  {journal}
  {\bibinfo  {journal} {arXiv e-prints}\ ,\ \bibinfo {eid} {cond-mat/0403055}}
  (\bibinfo {year} {2004})},\ \Eprint {https://arxiv.org/abs/cond-mat/0403055}
  {arXiv:cond-mat/0403055 [cond-mat.str-el]} \BibitemShut {NoStop}%
\bibitem [{\citenamefont {Sandvik}(1998)}]{Sandvik98}%
  \BibitemOpen
  \bibfield  {author} {\bibinfo {author} {\bibfnamefont {A.}~\bibnamefont
  {Sandvik}},\ }\href {https://doi.org/10.1103/PhysRevB.57.10287} {\bibfield
  {journal} {\bibinfo  {journal} {Phys. Rev. B}\ }\textbf {\bibinfo {volume}
  {57}},\ \bibinfo {pages} {10287} (\bibinfo {year} {1998})}\BibitemShut
  {NoStop}%
\bibitem [{\citenamefont {Shao}\ and\ \citenamefont
  {Sandvik}(2023)}]{SHAO20231}%
  \BibitemOpen
  \bibfield  {author} {\bibinfo {author} {\bibfnamefont {H.}~\bibnamefont
  {Shao}}\ and\ \bibinfo {author} {\bibfnamefont {A.~W.}\ \bibnamefont
  {Sandvik}},\ }\href
  {https://doi.org/https://doi.org/10.1016/j.physrep.2022.11.002} {\bibfield
  {journal} {\bibinfo  {journal} {Physics Reports}\ }\textbf {\bibinfo {volume}
  {1003}},\ \bibinfo {pages} {1} (\bibinfo {year} {2023})},\ \bibinfo {note}
  {progress on stochastic analytic continuation of quantum Monte Carlo
  data}\BibitemShut {NoStop}%
\bibitem [{\citenamefont {Deguchi}\ \emph {et~al.}(2000)\citenamefont
  {Deguchi}, \citenamefont {Essler}, \citenamefont {Göhmann}, \citenamefont
  {Klümper}, \citenamefont {Korepin},\ and\ \citenamefont
  {Kusakabe}}]{DEGUCHI2000197}%
  \BibitemOpen
  \bibfield  {author} {\bibinfo {author} {\bibfnamefont {T.}~\bibnamefont
  {Deguchi}}, \bibinfo {author} {\bibfnamefont {F.}~\bibnamefont {Essler}},
  \bibinfo {author} {\bibfnamefont {F.}~\bibnamefont {Göhmann}}, \bibinfo
  {author} {\bibfnamefont {A.}~\bibnamefont {Klümper}}, \bibinfo {author}
  {\bibfnamefont {V.}~\bibnamefont {Korepin}},\ and\ \bibinfo {author}
  {\bibfnamefont {K.}~\bibnamefont {Kusakabe}},\ }\href
  {https://doi.org/https://doi.org/10.1016/S0370-1573(00)00010-7} {\bibfield
  {journal} {\bibinfo  {journal} {Physics Reports}\ }\textbf {\bibinfo {volume}
  {331}},\ \bibinfo {pages} {197} (\bibinfo {year} {2000})}\BibitemShut
  {NoStop}%
\bibitem [{\citenamefont {Oka}\ and\ \citenamefont
  {Aoki}(2010)}]{oka2010dielectric}%
  \BibitemOpen
  \bibfield  {author} {\bibinfo {author} {\bibfnamefont {T.}~\bibnamefont
  {Oka}}\ and\ \bibinfo {author} {\bibfnamefont {H.}~\bibnamefont {Aoki}},\
  }\href@noop {} {\bibfield  {journal} {\bibinfo  {journal} {Physical Review
  B—Condensed Matter and Materials Physics}\ }\textbf {\bibinfo {volume}
  {81}},\ \bibinfo {pages} {033103} (\bibinfo {year} {2010})}\BibitemShut
  {NoStop}%
\end{thebibliography}%

\end{document}